\documentstyle[epsfig]{ioplppt}                
\begin{document}

\title{Strangeness production in high density baryon matter}
\author{R. Ganz}
\address{University of Illinois Chicago, USA\\ 
and \\
Max Planck Institut f\"ur Physik, \\ 
         F\"ohringer Ring 6, \\
         D-80805 M\"unchen, Germany}

\author{for the E917 collaboration}

\author{B.B.~Back$^{1}$, R.R.~Betts$^{1,6}$, H.C.~Britt$^{5}$, J.~Chang$^{3}$,
W.C.~Chang$^{3}$, C.Y.~Chi$^{4}$, Y.Y.~Chu$^{2}$, J.B.~Cumming$^{2}$,
J.C.~Dunlop$^{8}$, W.~Eldredge$^{3}$, S.Y.~Fung$^{3}$, R.~Ganz$^{6,9}$,
E.~Garcia$^{7}$, A.~Gillitzer$^{1,10}$, G.~Heintzelman$^{8}$,
W.F.~Henning$^{1}$, D.J.~Hofman$^{1}$, B.~Holzman$^{1,6}$, J.H.~Kang$^{12}$,
E.J.~Kim$^{12}$, S.Y.~Kim$^{12}$, Y.~Kwon$^{12}$, D.~McLeod$^{6}$,
A.~Mignerey$^{7}$, M.Moulson$^{4}$, V.~Nanal$^{1}$, C.~Ogilvie$^{8}$, R.~Pak$^{11}$,
A.~Ruangma$^{7}$, D.E.~Russ$^{7}$, R.K.~Seto$^{3}$, P.J.~Stanskas$^{7}$,
G.S.F.~Stephans$^{8}$, H.~Wang$^{3}$, F.L.H.~Wolfs$^{11}$, A.H.~Wuosmaa$^{1}$,
H.~Xiang$^{3}$, G.H.~Xu$^{3}$, H.B.~Yao$^{8}$, C.M.~Zou$^{3}$
}
\address{
$^{1}$ Argonne National Laboratory, Argonne, IL 60439 USA
\\
$^{2}$ Brookhaven National Laboratory, Chemistry Department, Upton, NY 11973 USA
\\
$^{3}$ University of California Riverside, Riverside, CA 92521 USA
\\
$^{4}$ Columbia University, Nevis Laboratories, Irvington, NY 10533 USA
\\
$^{5}$ Department of Energy, Division of Nuclear Physics, Germantown, MD 20874 USA
\\
$^{6}$ University of Illinois at Chicago, Chicago, IL 60607 USA
\\
$^{7}$ University of Maryland, College Park, MD 20742 USA
\\
$^{8}$ Massachusetts Institute of Technology, Cambridge, MA 02139 USA
\\
$^{9}$ Max Planck Institut f\"ur Physik, D-80805 M\"unchen,  Germany
\\
$^{10}$ Technische Universit\"at M\"unchen, D-85748 Garching, Germany
\\
$^{11}$ University of Rochester, Rochester, NY 14627 USA
\\
$^{12}$ Yonsei University, Seoul 120-749, South Korea
}

\begin{abstract}
Strangeness production in heavy-ion collisions, when compared to proton
proton collisions, is potentially a sensitive probe for collective energy deposition
and therefore for reaction mechanisms in general.
It may therefore provide insight into possible
QGP formation in dense nuclear matter.
To establish an understanding of the observed yields, 
a systematic study of high density baryon matter at different beam 
energies is essential.
This might also reveal possible discontinuities in the energy dependence
of the reaction mechanism. We present preliminary results for 
kaon production in Au+Au collisions at beam kinetic energies of 6, 8, and 10.7 GeV/u 
obtained by the E917 experiment at the AGS (BNL).
These measurements complement those carried out by the E866 
collaboration at 2, 4, and 10.7~GeV/u with a significantly 
enlarged data sample. In both experiments a large range of rapidities
was covered by taking data at different angular settings of the magnetic 
spectrometer.
\end{abstract}

\section{Motivation}
Collisions of heavy ions at energies around 10~AGeV
offer a unique opportunity to study baryonic matter at very 
high densities ($2\rho_0-8\rho_0$; where $\rho_0$ is the density of 
nuclei in their ground state).  
It has been proposed that such densities may result in the formation of
a new phase of matter, the Quark Gluon Plasma\cite{Ris}. But, even if the
existence of such a state remains highly speculative, a systematic
study of these collisions at various projectile energies and 
centralities (implying different densities) 
is important to gain insight into
reaction mechanisms and the behaviour of hadrons under such extreme 
conditions. It also provides 
a ``baseline''-measurement for comparisons
with experiments at higher center-of-mass energies (SPS, RHIC). It might place 
stringent constraints on models, thus yielding a 
more detailed understanding of relativistic heavy ion collisions in general.

The goal of the E917 experiment is to carry out such an investigation; 
it is achieved on one hand by a systematic variation of the initial state 
of the collision in terms of beam energy and centrality, and, on the other,
by a determination of a whole set of final state observables such as:
\begin{itemize}
\item Production yields, masses and mass-width of short-lived vector mesons ($\phi$, $K^*$) 
and baryonic resonances ($\Delta^{++}$),
\item cross sections for the production of $\bar{p}$ and strange hadrons,
\item high statistics two-particle intensity interferometry (HBT).
\end{itemize}
In this presentation we focus on the beam energy dependence of pion and kaon production 
at mid-rapidity.

\begin{figure}
\begin{center}
\mbox{\epsfig{file=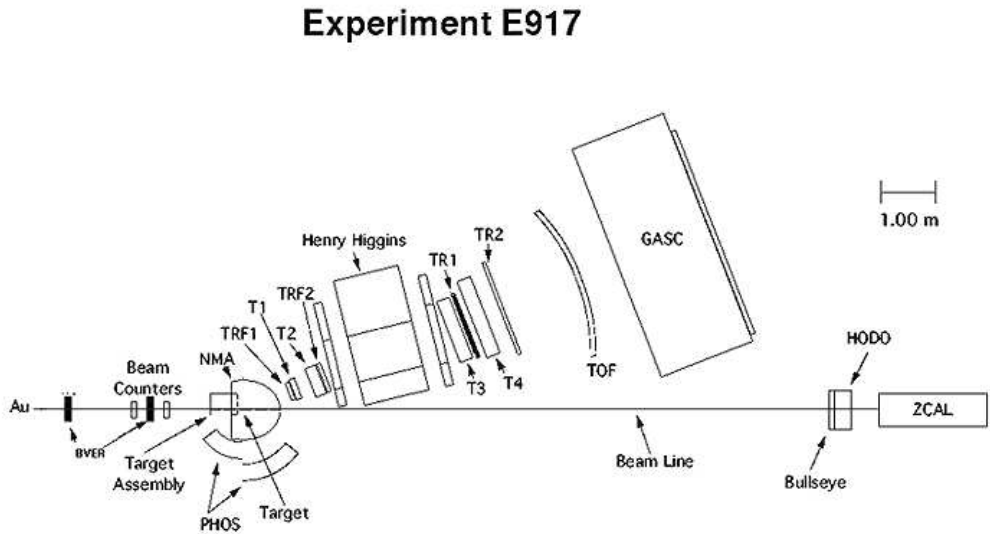,width=\linewidth}}                              
\end{center}
\caption%
{\small
A schematic view of the E917 spectrometer.}
\label{Fig:E917}
\end{figure}

\section{The E917 experiment}

\begin{figure}
\begin{center}
\mbox{\epsfig{file=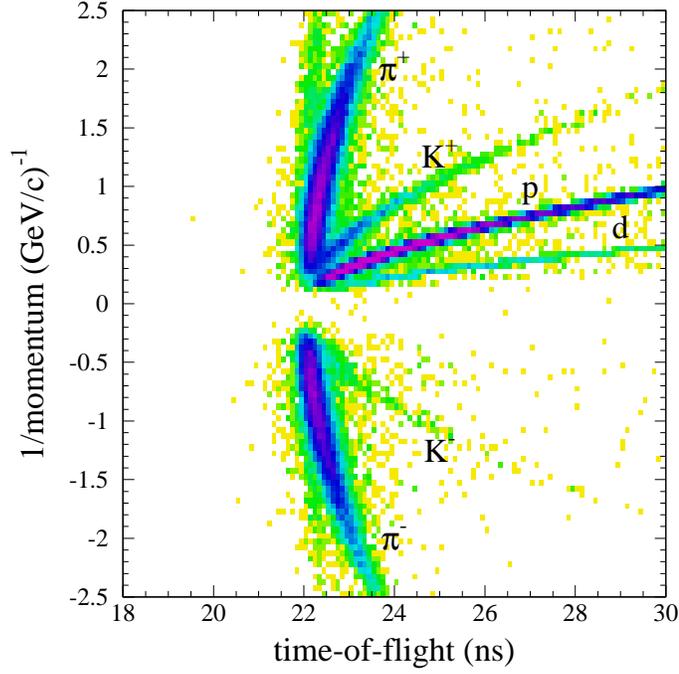,width=0.7\linewidth}}                              
\end{center}
\caption%
{\small
This figure demonstrates the particle identification capabilities of E917.
Plotted is the inverse of the particle momentum
extracted from a the drift chamber tracking versus the time-of-flight
measured in the TOF wall. The different branches are labeled according to
the different particle species.}
\label{Fig:PID}
\end{figure}

E917 is located at the AGS accelerator facility 
at Brookhaven National Laboratory, New York (BNL).
This experiment concludes a series of preceding 
experiments, E802, E859 and E866 \cite{Aki,Ahl,Shi,Vid}, 
which carried out various measurements with O, S, and Au beams.
In particular, E866  studied Au+Au collisions 
at 2, 4, and 10.7 AGeV projectile kinetic energies, 
which has been complemented by the E917 measurements
at 6, 8, and 10.7~AGeV with an upgraded experimental setup  
(figure~\ref{Fig:E917}; a detailed description of the E802 apparatus can be found in \cite{Abb}).
The main part of the E917 apparatus is a movable spectrometer arm,
which consists of the ``Henry-Higgins'' (HH) 
magnet ($\pm$~0.4~Tesla) located between sets of multi-wire drift chambers 
(T1-T4 and TRF1-TRF2), each comprising several wire planes with 
different orientations.
This allows for momentum measurement with
a resolution of $\Delta p/p\approx 1\%$\footnote{A resolution of 2\% was achieved
in the low energy measurement with a $\pm$0.2~Tesla HH field.}.  
In combination with
a time-of-flight scintillator wall (TOF; timing resolution $\delta t\approx 130$~ps), 
particle identification is achieved as shown in figure~\ref{Fig:PID}. 
The whole spectrometer can be rotated about the target and centered at polar angles 
in the range of 14$^o$-44$^o$ in 5$^o$ steps with respect to the beam axis; 
measurements over a wide range of rapidities are therefore possible. 
At each setting, a solid angle of 25~msr is covered by the spectrometer.

A set of beam counters detects and validates incoming beam particles, 
vetos upstream interactions 
and defines a ``time-zero'' for the time-of-flight measurement. 
An upper limit on the signal from the ``Bullseye'' detector (BE) 
-- being insensitive on the 25\% of most peripheral collision --
is used to trigger on interactions in the 4\% interaction length gold target. 
To enrich the data sample with events with a specific combination of identified particles
(eg. 2K/$\bar{p}$), a Level 2 trigger has been implemented, based on 
characteristic hit patterns in the TOF system and two 
multi-wire proportional chambers (TR1-TR2). 

To characterize the centrality of a collision two methods were used.
One utilized the multiplicity measured in the New Multiplicity Array (NMA), 
which surrounds the target area. The other allows for an estimate of  
the number of participants in the
collision via the energy ($E_{\mathrm{ZCAL}}$) 
recorded in the Zero Degree Calorimeter 
(ZCAL; 1.5$^o$ opening angle) and the projectile energy ($E_{\mathrm{Proj}}$) from 
$N_{\mathrm{part}}\approx 197\times(1-E_{\mathrm{ZCAL}}/E_{\mathrm{Proj.}})$.

An important new feature of the E917 apparatus is an event-by-event 
tracking of beam particles 
by four position sensitive scintillating fiber detectors 
(BVER; $\delta x\approx 200\mu$m) mounted upstream of the target \cite{BBB}. 
In combination with the downstream scintillator wall (HODO) this allows for an 
event-by-event reconstruction of the reaction plane of non-central collisions.
Further recent improvements of the E917 setup are improvements to
the TR1 trigger chamber and a more reliable data acquisition system.

\section{Preliminary results}

Preliminary results from an analysis
of a subset of the Au+Au data taken in November '96- January '97 
are presented in the following sections 
in comparison with results obtained by the 
E866 collaboration\cite{Dun,Ogi2,Set,Ogi}.
 
\begin{figure}
\begin{center}
\mbox{\epsfig{file=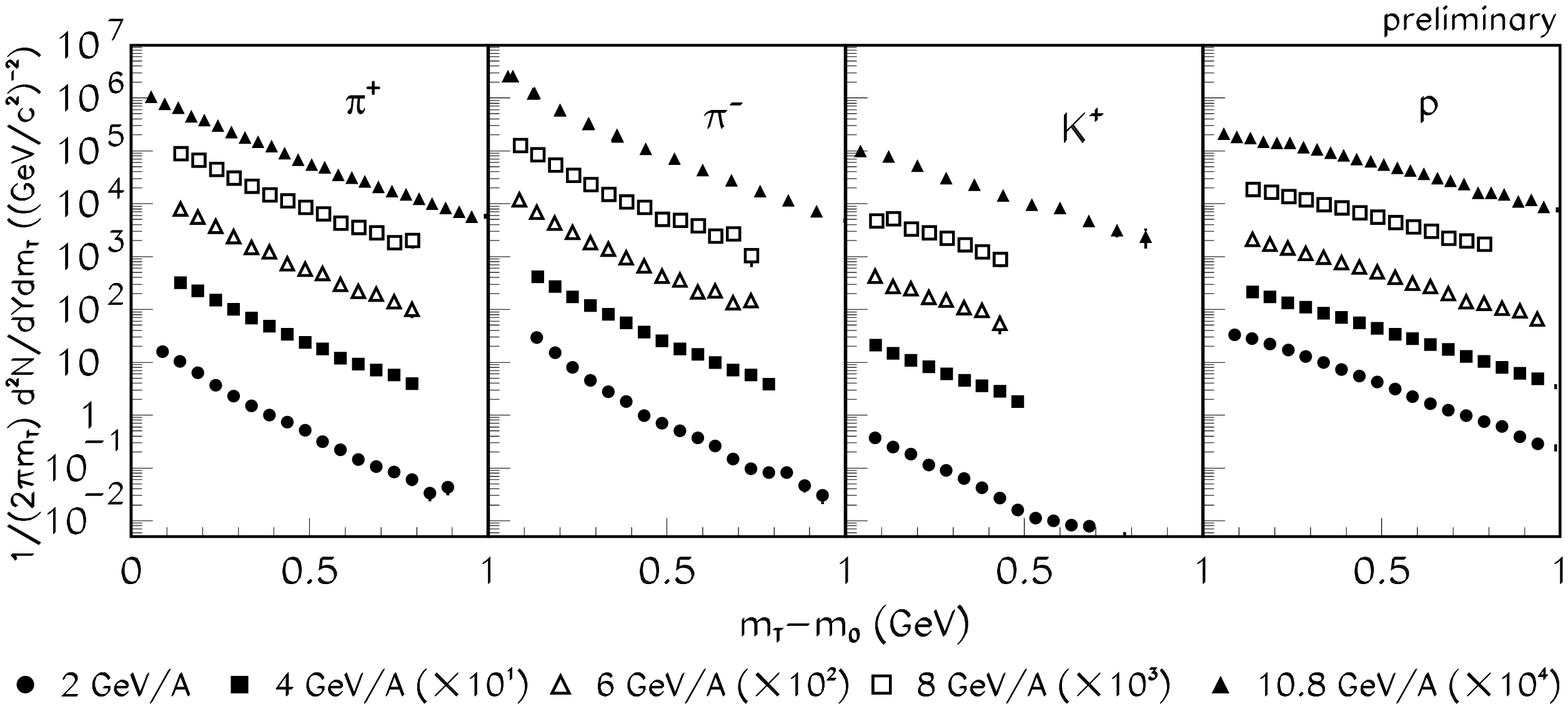,width=1.05\linewidth}}               
\end{center}
\caption%
{\small
Invariant particle yields are plotted on a logarithmic scale 
vs. $m_{\mathrm{t}}-m_0$ at $\Delta y\pm 0.25$ around mid-rapidity
for central Au+Au collision. The yields for beam kinetic energies 4~AGeV and higher are 
scaled as labeled. The errors bars shown are statistical only.}
\label{Fig:Yieldmt}
\end{figure}

Figure~\ref{Fig:Yieldmt} shows the invariant yields 
at mid-rapidity $\mid y-y_{\mathrm{cm}}\mid/y_{\mathrm{cm}}<0.25$
of pions, kaons, and protons 
as function of $m_{\mathrm{t}}=\sqrt{p_{\mathrm{t}}^2+m_0^2}$ for beam kinetic energies 
of 2, 4, 6, 8, and 10.7~AGeV
The most central 8\% 
of the total interaction cross section of 6.8~b \cite{Wad} have been selected.
The error bars are statistical only. The systematic error
varies betweeen 5\% - 20 \% dependent on particle type.
 
In case of kaons a single exponential describes the spectral shape of the yield quite 
well, whereas for the pions a double exponential is better suited due to the enhancement 
of the yield at low transverse momenta, which is most pronounced for $\pi^-$.
In the case of protons the yield follows a Boltzmann shape as a function of $<m_{\mathrm{t}}>$.
 
\begin{figure}
\begin{center}
\mbox{\epsfig{file=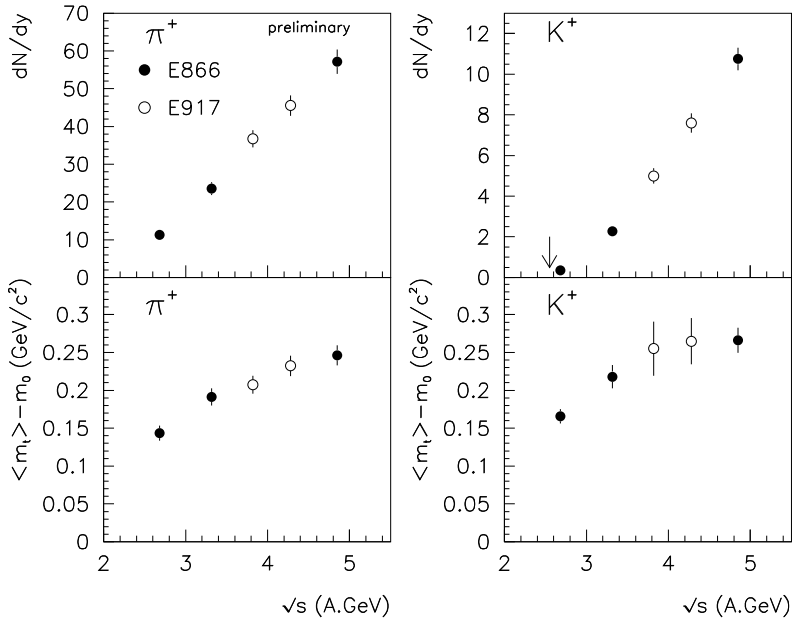,width=0.8\linewidth}}               
\end{center}
\caption%
{\small
The upper two panels show the invariant yields for K$^+$ and $\pi^+$ 
integrated over $m_{\mathrm{t}}$ versus the center-of mass energy of the collision. 
The arrow in the upper right panel indicates the 
pp$\rightarrow\bar{\Lambda}$K$^+$p threshold.
The two lower panels show the corresponding dependence of the 
average transverse mass. The error bars include statistical errors and 5\% point-to-point
systematic uncertainty.}
\label{Fig:Yield}
\end{figure}

Instead of extracting slope parameters (``temperature'') from the 
shapes, the average value $<m_{\mathrm{t}}-m_0>$ might be a better way to characterize the amount of 
longitudinal kinetic energy of the projectile 
transferred into ``thermal'' motion\footnote{Which part of the transverse motion 
is due to thermal motion and which due to collective phenomena like expansion 
is currently under investigation.} of final state particles.
Figure~\ref{Fig:Yield} compares
the energy dependence of $<m_{\mathrm{t}}-m_0>$ for $\pi^+$ and K$^+$ 
with the energy dependence of total yields $dN/dy$
at mid-rapidity. 
The plotted values are derived from extrapolating the measured
$m_{\mathrm{t}}$-spectrum using
the functional forms mentioned above. 
With increasing available energy $(\sqrt{s})$ both observables rise smoothly and no 
discontinuities are found, as might naively be expected if a 
transition to a QGP had occured.
The figure also shows that $<m_{\mathrm{t}}>$ increases rather slowly
over the energy range considered, and show a hint of saturation
at the higher end. 
The number of produced particles, $dN/dy$,
increases more rapidly, not only for the 
kaons produced close to threshold, but 
also for the pions. Taking into account 
that stopping seems to be rather complete at AGS energies \cite{Vid},
this observation means that the extra available energy 
preferentially goes into particle production 
and not into an increase of their transverse energy.

\begin{figure}
\begin{center}
\mbox{\epsfig{file=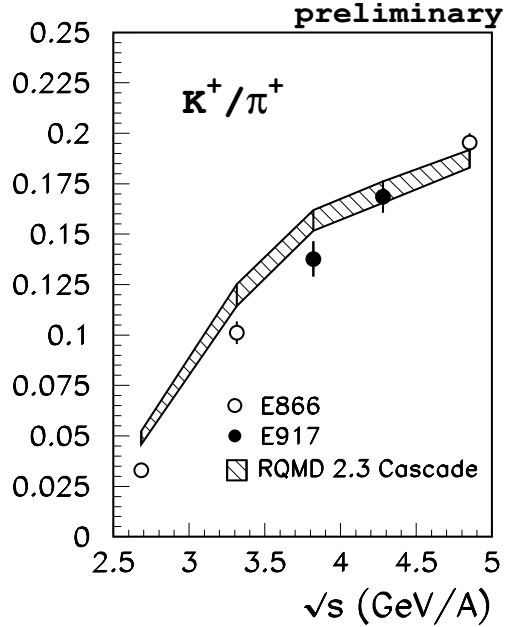,width=0.7\linewidth}}                              
\end{center}
\caption%
{\small
The ratio of K$^+$- to $\pi^+$- yield at mid-rapidity plotted 
versus the available energy for central Au+Au collisions. 
Errors are statistical only. The hatched area correspond to 
predictions from the RQMD cascade model.} 
\label{Fig:RQMD}
\end{figure}

The fact that this increase with $\sqrt{s}$ is more pronounced for the kaon yields 
is expressed in terms of the ratio K$^+/\pi^+$, 
which 
rises continuously with increasing available energy (figure~\ref{Fig:RQMD}). Such a behaviour
might be expected considering that the threshold for 
pp$\rightarrow\bar{\Lambda}\mathrm{K}^+$p is at $\sqrt{s}=2.343$~AGeV\cite{Cos}. 
Moreover, the observed energy dependence is in good agreement with predictions from the 
RQMD cascade model\cite{Sor}.

Further up the energy scale, a K$^+/\pi^+$-ratio of 0.145 
was deduced from results of the NA49 measurement\cite{Sey,Bor}
of Pb+Pb collisions at $\sqrt{s}=17.2$~AGeV\footnote{The 
reference gives the ratio $(\mathrm{K}+\bar{\mathrm{K}})/\pi=0.145$ 
extrapolated to $4\pi$, from which
the K$^+/\pi^+\approx0.14$ in the text is estimated by using 
$\mathrm{K}^0_s\approx\mathrm{K}^+$, $\mathrm{K}^+/\mathrm{K}^-\approx 1.8$, 
and $\pi^-\approx\pi^0\approx\pi^+$.}.
This is significantly below what is found in the measurement at 
the highest AGS energy and prompts the question, at which energy  
the maximum is reached and whether this might 
originate from a change of the kaon production mechanism. 
Further investigations, especially in the intermediate energy regime 
as proposed by NA49\cite{Pro}, might help to resolve this puzzle.

\begin{figure}
\begin{center}
\mbox{\epsfig{file=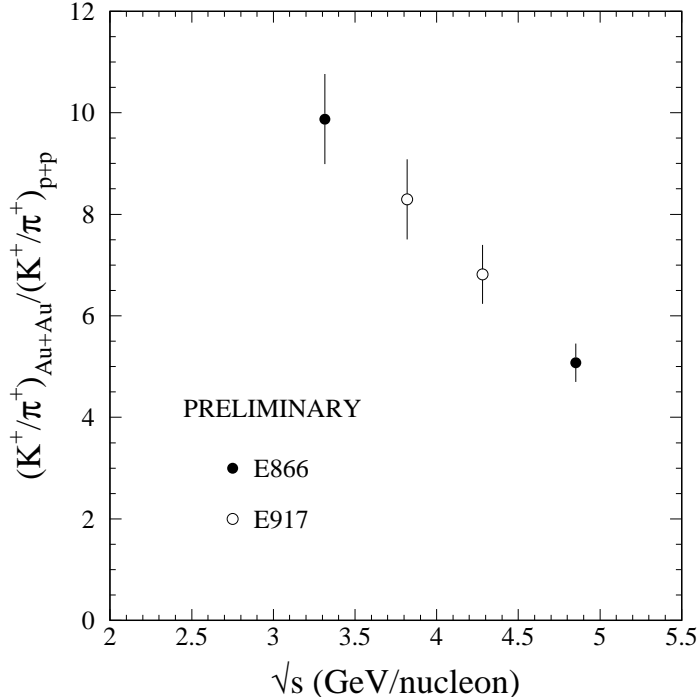,width=0.7\linewidth}}                              
\end{center}
\caption%
{\small
The ratio K$^+$ to $\pi^+$ measured in Au+Au collisions is 
devided by the corresponding value of p+p collisions.
The ratio in Au+Au is determined at mid rapidity, whereas the  
p-p data in the literature is integrated over all rapidities.}
\label{Fig:Double}
\end{figure}

Strangeness enhancement is one of the potential signatures at the transition to 
the Quark Gluon Plasma. One way to quantify ``enhancement'' is to normalize the 
K$^+/\pi^+$-ratio measured in Au+Au collisions to the one observed in 
p+p collisions at the same center-of-mass energy. 
Such a double ratio, if larger than one, suggests that
we are not dealing simply with a superposition of ``A times'' 
a nucleon-nucleon collision, but that we must consider effects such as multiple collisions,
collective phenomena, thermalization and even the existence 
of a QGP phase, all of which 
may contribute to an enhanced production of kaons, but which should be absent 
in proton-proton collisions. 
For the Au+Au data this double ratio is shown in figure~\ref{Fig:Double}.
It should be noted that the ratio for Au+Au collisions is derived at mid-rapidity,
whereas the ratio for p+p collisions taken from literature \cite{Blo,Ros,Fed}
is given as the integral over all rapidities.
The data point at the lowest beam kinetic energy (2~AGeV) has been omitted because of a large 
uncertainty of the p+p kaon yield. No correction for the Fermi motion or isospin 
difference between Au+Au and p+p has been applied.

This double ratio grows when getting closer to the kaon production threshold in pp
collisions, which underlines the importance of secondary collisions for production 
of kaons from nucleus-nucleus collisions in this energy regime.  
Again, a comparison to the $\sqrt{s}=17.2$~AGeV Pb+Pb measurement is instructive. Normalized to 
p+p data\cite{Gad,Agu}
at $\sqrt{s}=19.4$~AGeV\footnote{The kaon yield in p+p collisions
is almost constant ($\approx 0.087$) for center-of-mass energies of 10~GeV--100~GeV.}
the double ratio $($K$^+/\pi^+)_{AuAu}/($K$^+/\pi^+)_{pp}\approx 1.7$ still significantly 
exceeds unity. 
Even if threshold effects, which 
dominate the AGS energy regime, are negligible, the enhancement still persists, as
different channels for kaon production may come in to play.

\section{Summary and conclusion}

E917 has carried  out a study of particle production by systematic variation
of the initial state of the collision in terms of beam energy, centrality and 
reaction plane. Results on the energy dependence of the kaon and the pion yield 
from a preliminary E917 analysis in conjunction with
a set of similar measurements from the E866 collaboration
do not exhibit any sudden changes indicating new phenomena.
The increase of available energy appears predominantly as an increase of the 
number of particles and not as an increase of transverse momentum.
The comparison of the K$^+/\pi^+$ ratio to results at CERN SPS indicates 
a maximum
at intermediate energy, which might indicate a change of production mechanism.
This ratio normalized to p+p collisions, seems to level off at higher energies, 
which is also not fully understood. A more detailed analysis of the complete data set 
will shed light one these questions.

\end{document}